\documentclass[preprint2]{aastex6}

\usepackage{hyperref}
\usepackage{comment}
\hypersetup{citecolor = blue}

\fullcollaborationName{The Friends of AASTeX Collaboration}

\begin{document}


\title{Constraining CIII] Emission in a sample of five luminous \textit{z} = 5.7 galaxies}


\author{Jiani Ding\altaffilmark{1}, Zheng Cai\altaffilmark{2,7}$^{,\ast}$, Xiaohui Fan\altaffilmark{1}, Daniel P. Stark\altaffilmark{1}, Fuyan Bian\altaffilmark{3}, Linhua Jiang\altaffilmark{4}, Ian D. McGreer\altaffilmark{1}, Brant E. Robertson\altaffilmark{5}, Brian Siana\altaffilmark{6}}

\altaffiltext{1}{Steward Observatory, University of Arizona, 933 N Cherry Ave., Tucson, AZ, 85721, USA}
\altaffiltext{2}{UCO/Lick Observatory, University of California, 1156 High Street, Santa Cruz, CA, 95064, USA}
\altaffiltext{3}{Research School of Astronomy and Astrophysics, The Australian National University, Cotter Road, Weston Creek, ACT 2611, Australia}
\altaffiltext{4}{Kavli Institute for Astronomy and Astrophysics, Peking University, Beijing 100871, China}
\altaffiltext{5}{Department of Astronomy and Astrophysics, UC Santa Cruz, CA 95064, USA}
\altaffiltext{6}{Dept of Physics and Astronomy, UC Riverside, 900 University Ave, Riverside, CA, 92521, USA}
\altaffiltext{7}{Hubble Fellow \newline $\ast$ zcai@ucolick.org}

\begin{abstract}

Recent observations have suggested that the CIII]$\lambda1907/1909$ emission lines could be alternative diagnostic lines for galaxies in the reionization epoch. We use the F128N narrowband filter on the Hubble Space Telescope's ($\it{HST}$) Wide Field Camera 3 (WFC3) to search for CIII] emission in a sample of five galaxies at \textit{z} = 5.7 in the Subaru Deep Field and the Subaru/XMM-Newton Deep Field. Using the F128N narrowband imaging, together with the broadband imaging, we do not detect CIII] emission for the five galaxies with $J_{\rm{AB}}$ ranging from 24.10 -- 27.00 in our sample. For the brightest galaxy J132416.13+274411.6 in our sample (\textit{z} = 5.70, $J_{\rm{AB}} = 24.10$), which has a significantly higher signal to noise, we report a CIII] flux of  $3.34\pm1.81 \times 10^{-18}$ $\mathrm{erg\ s^{-1}\ cm^{-2}}$, which places a stringent 3-$\rm\sigma$ upper limit of $5.43\times 10^{-18}$ $\mathrm{erg\ s^{-1}\ cm^{-2}}$ on CIII] flux and 6.57 \AA\ on the CIII] equivalent width. Using the stacked image, we put a 3-$\rm\sigma$ upper limit on the mean CIII] flux of $\mathrm{2.55\times10^{-18}\ erg\ s^{-1}\ cm^{-2}}$, and a  3-$\rm\sigma$ upper limit on the mean CIII]  equivalent width of 4.20 $\mathrm{\AA}$ for this sample of galaxies at \textit{z} = 5.70. Combined with strong CIII] detection reported among high-\textit{z} galaxies in the literature, our observations suggest that the equivalent widths of CIII] from galaxies at \textit{z} $>$ 5.70 exhibit a wide range of distribution.  Our strong limits on CIII] emission could be used as a guide for future observations in the reionization epoch. 


\end{abstract}
 



\section{Introduction} \label{sec:intro}
\setlength{\parindent}{1em}
In recent years, we have witnessed great progress in studying galaxies at \textit{z} $>$ 6. Several deep field surveys using the Hubble Space Telescope ($\it{HST}$) and largest ground-based telescopes (\citealt{Bouwens2015b,Ellis2013,Finkelstein2015,McLure2013}; for a review see \citealt{Stark2016}) have given rise to hundreds of \textit{z} $>$ 6 galaxy candidates.  One of the main challenges for redshift confirmation in \textit{z} $>$ 6 galaxy candidates is the increasing attenuation of Ly$\alpha$ emission due to a rising intergalactic medium (IGM) neutral fraction in the end of the reionization epoch  \citep{Planck2016,Caruana2014,Choudhury2015,Mesinger2015}. As a result, the efficiency and validity of using Ly$\alpha$ emission to study \textit{z} $>$ 6 galaxies are limited, leading to the search for an alternative diagnostic line. 
Spectroscopic surveys among star forming galaxies at \textit{z} $\sim$ 2 \citep{Stark2014} and \textit{z} $\sim$ 7 \citep{Stark2015,Stark2017} reveal that CIII]$\lambda1907/1909$ could be used as alternative emission lines for determining the redshifts for high-$z$ galaixes.

 The strength of various nebular emission lines, such as HeII $\lambda1640$, CIV$\lambda1549$, OIII]$\lambda1661/1667$, and CIII]$\lambda1907/1909$, can help to constrain galaxy physical properties, including the ionization parameters and metallicities \cite[e.g.,][]{Erb2010,Cai2011}. Recently, several efforts have been conducted to investigate the CIII] emission from high-$z$ galaxies \cite[e.g.,][]{Stark2014,Stark2015,Stark2017,Zitrin2015,Rigby2015}. The CIII] emission has been detected in a sample of Lyman Break Galaxies (LBGs) at \textit{z} $\sim$ 3 \citep{Shapley2003} and in a sample of low mass, low luminosity star forming galaxies at \textit{z} $\sim$ 2 \citep{Stark2014}. Previous studies \citep{Stark2014,Jaskot2016} suggest that galaxies with strong CIII] emission should have high ionization parameters ($\mathrm{log U}$ varies from -2.16 to -1.84), low metallicities ($\mathrm{0.04-0.13\ Z_{\odot}}$) and younger stellar populations ($\mathrm{6-50\ Myr}$) (detailed ranges from \cite{Stark2014}). The photoionization models from \cite{Jaskot2016} support that galaxies with strong CIII] emissions  may have $\mathrm{log U} \ge -2$ and metallicity Z $\sim \mathrm{0.14\ Z_{\odot}}$ after taking into account the decline of C/O ratio with decreasing metallicity. Such conditions may be common in typical $z\gtrsim$ 6 galaxies \citep{Stark2014,Jaskot2016}. \citet{Stark2015, Stark2017} conducted a pilot survey on the CIII] emissions for $z\gtrsim6-7$ galaxies which are expected to have strong [OIII] and $\rm{H{\beta}}$ emission from the rest-frame optical photometric measurements.  They successfully detected robust CIII] emission lines in one galaxy. Nevertheless, the current sample sizes in terms of galaxy numbers and galaxy properties are still too small for conducting a robust statistical study on CIII] properties at $z\gtrsim6$.

 To better understand the fraction of galaxies that host strong CIII] lines in the reionization epoch,  it is crucial to conduct a systematic study to search for  CIII] emission in a statistical sample of high-redshift galaxies. There have been $> 100$ high-redshift Ly$\alpha$ emitters discovered at $z \sim 5.7$. The most extensive surveys are those using deep imaging in the Subaru Deep Field (SDF, \cite{Kashikawa2011}) and Subaru/XMM-Newton Deep Field (SXDF, \cite{Ouchi2005}).  The SDF field contains the brightest Ly$\alpha$ emitter (LAE), J132416.13+274411.6, at \textit{z} = 5.70 \citep{Shimasaku2006}, and the SXDF field contains an overdensity of LAEs at $z\approx5.70$ \citep{Ouchi2005}. Furthermore, deep $\it{HST}$ observations in $Y$, $J$, and $H$ bands have been conducted to precisely measure continuum levels and rest-frame ultraviolet (UV) morphologies of the LAEs in these two fields \citep{Jiang2013}.  The LAEs at $z\approx5.7$ in SDF and SXDF are ideal for conducting a systematic study of CIII] emission.   
 
\indent In this paper, we present our deep, high-resolution narrowband (F128N) and broadband imaging (F125W) in SDF and SXDF fields using the $\it{HST}$ Wide Field Camera 3 (WFC3) imaging. At $z=5.7$, the CIII] emission is redshifted to the the most sensitive part of the F128 narrowband filter. Our $\it{HST}$ observations targeted five \textit{z} = 5.7 galaxies in CIII] emission using two $\it{HST}$ pointings. The high sensitivity and low sky background with WFC3 allow us to probe CIII] lines in LAEs at a level that is difficult to achieve from the ground.  In this paper, we report a tentative 2-$\sigma$ flux excess in the F128N filter for the brightest LAE at $z=5.7$ in the SDF field. For the other four galaxies in the SXDF protocluster,  we report a null detection of CIII] emission. Using this sample of galaxies, we place a stringent upper limit on the CIII] fluxes and equivalent widths (EW) for the five galaxies in our sample at $z=5.7$.  Organization of this letter is as follows. In \S2, we discuss our $\it{HST}$ observation and data reduction. In \S3, we measure the UV slope, continuum, and morphology for the brightest galaxy in SDF and report 3-$\rm\sigma$ upper limits for CIII]$\lambda1907/1909$ emission line fluxes in the other four galaxies in the SXDF.
 In \S4, we discuss the astrophysical implications of our photometry results. Throughout the whole paper, we adopt a flat lambda-CDM cosmology with $\rm{\Omega_{\lambda}=0.7,\ \Omega_{m}=0.3\ and\ H_{0}=70\ km\ s^{-1}\ Mpc^{-1}}$.


\section{OBSERVATION AND DATA REDUCTION} \label{sec:style}
 
We conduct a deep $\it{HST}$/WFC3 narrowband (F128N) and broadband (F125W) imaging in a sample of five galaxies. These galaxies are LAEs at \textit{z} = 5.7 selected from the SDF \cite{Kashikawa2011} and the SXDF \cite{Ouchi2005}. The galaxies described in this paper are listed in Table \ref{tab:sxdf}. The redshift range of our sample is \textit{z} = 5.69 -- 5.75. The redshifts are determined based on their Ly$\alpha$ emission. From \cite{Erb2014}, the average Ly$\alpha$ velocity offset of LAEs with the absolute magnitudes of $\rm{M_{UV} \le -19.3}$ (similar to that of our sample) at \textit{z} $\sim$ 2 -- 3 is $\approx 300$ $\rm{km\ s^{-1}}$. Assuming the Ly$\alpha$ systematic velocity offset for our LAE sample is 300 km s$^{-1}$, the CIII]$\lambda$1907/1909 emission of our sample galaxies resides at 12745.1 $\mathrm{\AA}$ -- 12872.9 $\mathrm{\AA}$, securely lying on most of the sensitive wavelengths of the F128N filter (${\lambda{c}=12832\ \rm{\AA}, FWHM = 159\ \rm{\AA}}$) we used.

In Cycle-22, we use 14 orbits to conduct the $\it{HST}$/\ WFC3 observations in both SDF and SXDF fields. To detect the CIII] line for brightest LAE at $z=5.7$,  J132416.13+274411.6 (source 1, see Table \ref{tab:sxdf}), we use six-orbits ($\sim$ 16000 seconds integration time) F128N imaging in the SDF field. Deep $\it{HST}$ F110W and F160W broadband imaging of the brightest galaxy (source 1) have already been conducted in \cite{Jiang2013}, and we use the F110W and F160W imaging to determine the continuum level and continuum slope of source 1. 
We assign another six orbits ($\sim$16000 seconds integration time) F128N imaging for the other four galaxies at \textit{z} = 5.7 in the SXDF (sources 2 -- 5, see Table \ref{tab:sxdf}).  To conduct the accurate continuum subtraction for our galaxies in the SXDF, we use two orbits (5223.5 second integration time) F125W imaging for sources 2 - 5 in the SXDF. These observations allow us to reach a F128N depth of  1-$\sigma$  = $1\times10^{-18}\  \mathrm{erg\ s^{-1}\ cm^{-2}}$ in the SDF and SXDF fields, enabling the detection of CIII] emission as weak as $\sim$ 6 \AA\ at 3-$\sigma$ level for the brightest target in our sample.

We distribute our entire observations into four individual visits.  For each visit, a standard four-point dither sequence is applied to populate each of the two orbits. The data reduction is conducted using Multidrizzle \citep{Koekemoer2002}, and the detailed procedures follow the descriptions in \citet{Cai2011,Cai2015}. To optimize the output data quality, we choose a final output pixel scale of 0.06$"$ instead of the initial pixel scale 0.13$"$ and final pixfac parameter 0.7 (shrinking pixel area) after different trials of combinations of parameters. The final output images we obtained from different $\it{HST}$ band imaging for source 1 is showed in Figure \ref{fig:figuresdf} and for sources 2 -- 5 are showed in Figure \ref{fig:figure2}.


\section{MEASURED CIII] FLUXES AND UPPER LIMITS} \label{sec:floats}

\begin{figure*}[ht!]
\centering

\includegraphics[width=150mm]{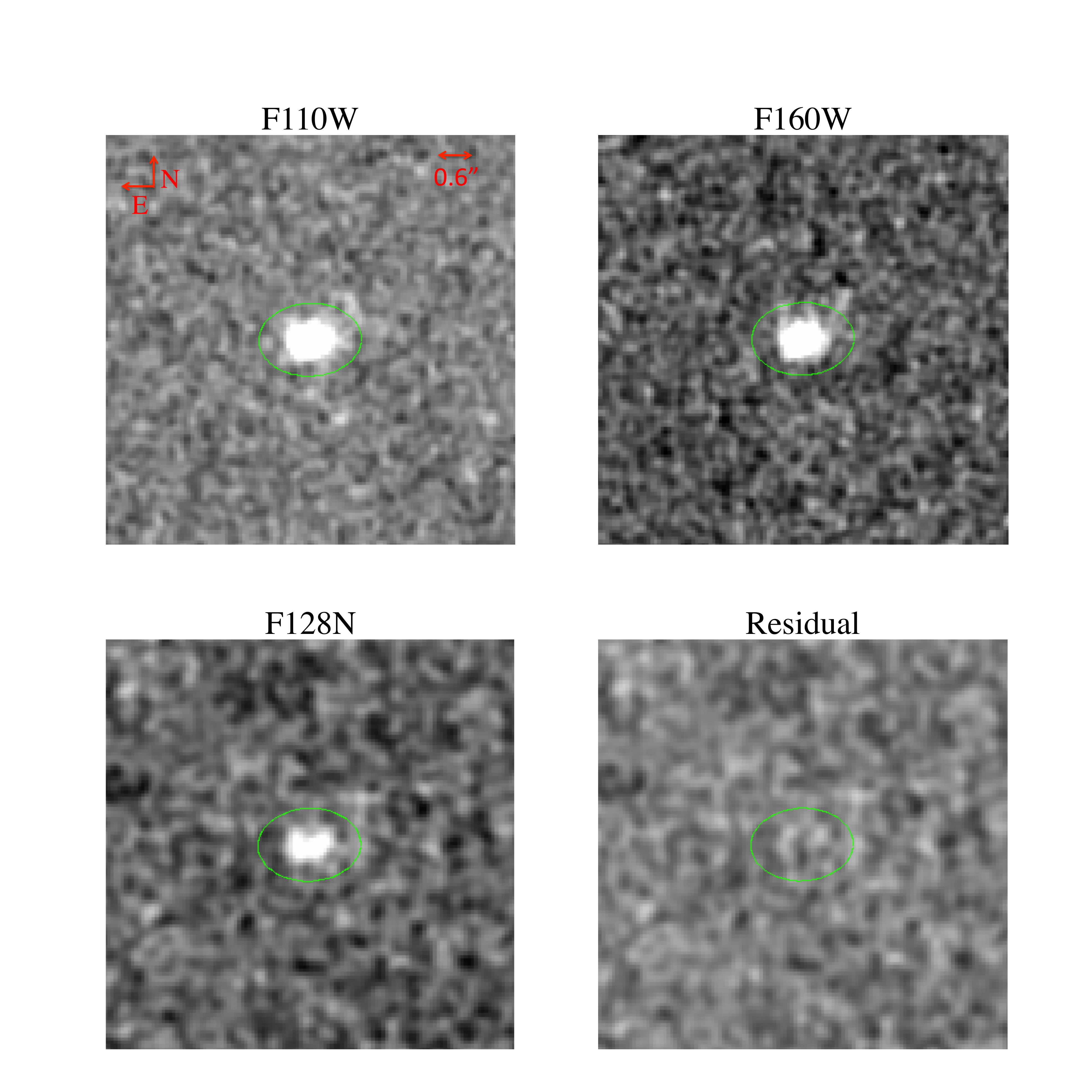}
\epsscale{1.5}
\caption{High-resolution smoothed F110W broadband image (top left), F160W broadband image (top right), F128N narrowband image (bottom left), and subtracted image (bottom right) for brightest galaxy in Subaru deep field (Source 1) with apertures of 1.8 kron factor and 2.5 semi-major radius. Based on the photometry result, we found a 2-$\rm\sigma$ flux excess of CIII] emission in source 1, which places a 3-$\rm{\sigma}$ upper limit of 6.57 \AA\  on the CIII] equivalent width for this galaxy. All graphs have the same scale and orientation. }\label{fig:figuresdf}
\end{figure*}

\begin{figure*}[ht!]
\centering

\includegraphics[width=180mm]{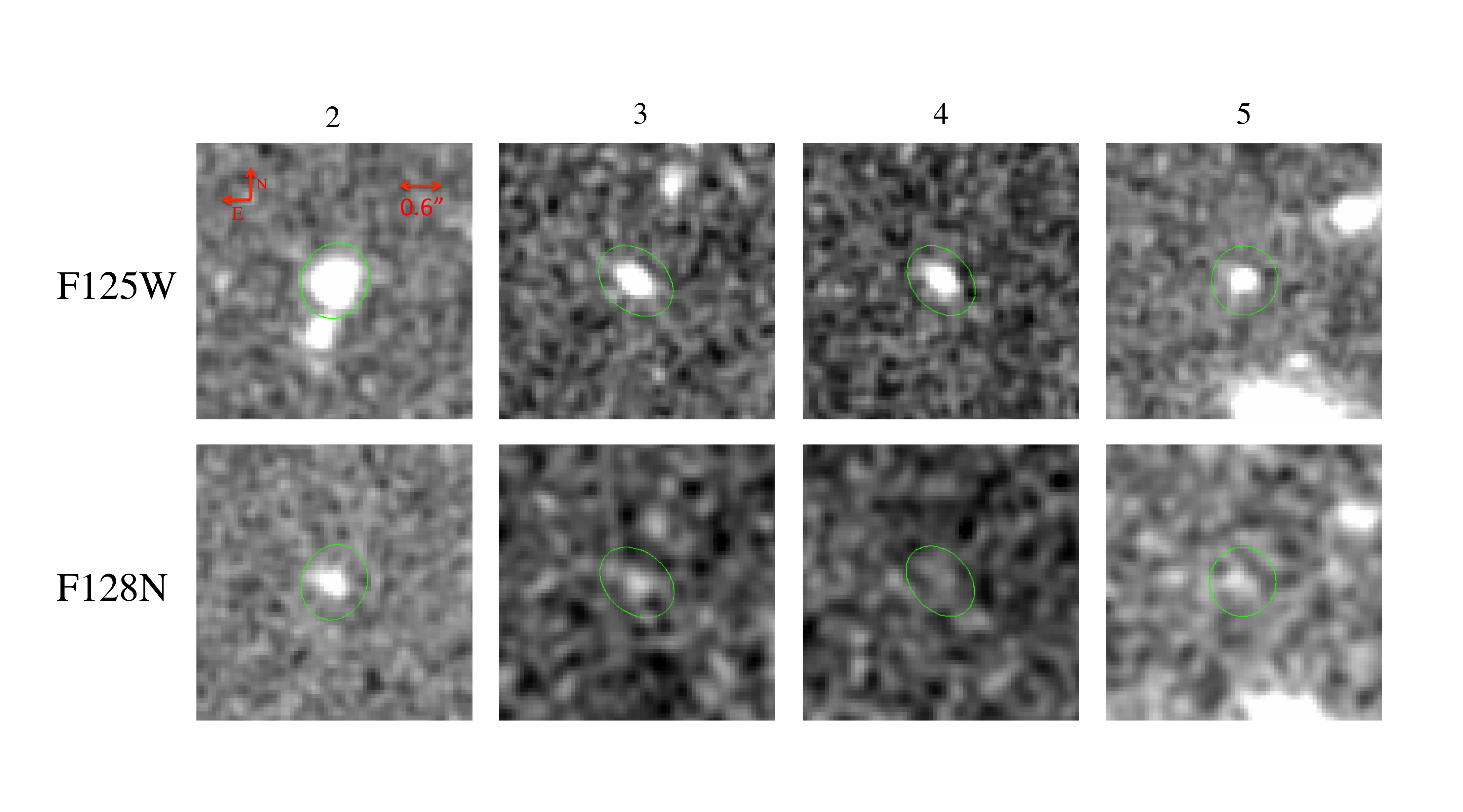}
\caption{Smoothed narrowband F128N images for source 2 to source 5 (bottom four images) and F125W broadband images for sources 2 to 5 (top four images) for galaxies (see Table \ref{tab:sxdf}) in Subaru/XMM-Newton deep field with elliptical apertures of 1.8 kron factor and 2.5 semi-major radius. Photometry results indicate no CIII] emission in these four galaxies. All graphs have the same scale and orientation. }\label{fig:figure2}
\end{figure*}

Similar to \citet{Cai2011,Cai2015}, we use SExtractor \citep{Bertin1996} to measure the fluxes in the narrowband (F128N) and broadband filters (F110W and F160W) for our galaxy sample. For source 1 -- 5, we apply an $\rm{MAG_{AUTO}}$ elliptical aperture with a kron factor of 1.8 and a minimum aperture of 2.5 semi-major radius to measure the flux in the the narrowband imaging. We apply the same aperture to the broadband imaging (Figure \ref{fig:figuresdf} and Figure \ref{fig:figure2}). 

\begin{figure*}[ht!]
\centering

\includegraphics[width=150mm]{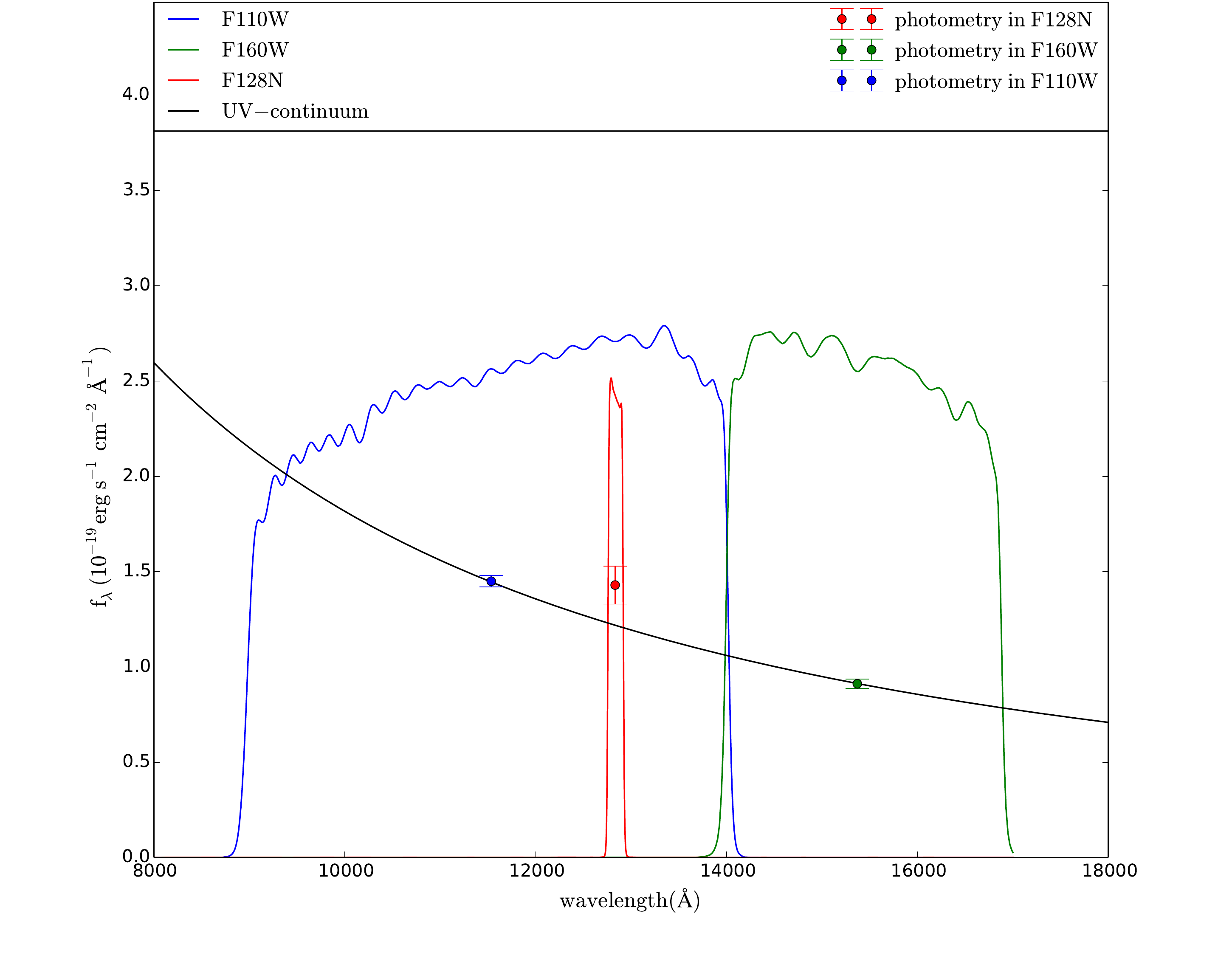}
\epsscale{1.5}
\caption{CIII] emission and the best fit spectrum (black line) with UV-continuum fitted by photometry in WFC3/F110W (blue dot with error bar) and photometry in WFC3/F160W (green dot with error bar) of the brightest galaxy in our sample in SDF. The photometry in WFC3/F128N (red dot with error bar) shows a 2-$\sigma$ excess of the WFC3/F128N flux compared with the fitted UV-continuum at the same wavelength. The filter response curves are plotted for WFC3/F110W filter (blue curve), WFC3/F160W (green curve) and WFC3/F128N filter (red curve). The photometry data from F110W and F160W bands is from \cite{Jiang2013}, while the photometry data from F128N is from our $\it{HST}$ observation.}\label{fig:figure3}
\end{figure*}

The photometry for source 1 is presented in Table \ref{tab:sxdf} and Figure \ref{fig:figure3}.  We use our photometry results in the F110W broadband and F160W broadband to fit the spectral energy distribution (SED), assuming a standard power law continuum $f_{\rm{C}}$ =  $ {\alpha(\lambda/10000{\rm{\AA})}}^{\beta}$, where $\alpha$ is a constant. We find $f_{\rm{C}}(\lambda) = 1.82\pm0.10 \times 10^{-19} \mathrm{ (\lambda/10000{\AA})}^{-1.60\pm0.20}$ $\mathrm{erg\ s^{-1}\ cm^{-2} \AA^{-1}}$. The F128N filter is included in the F110W filter bandpass. In this calculation, we assume the F110W flux density as a pure continuum (contribution from CIII] emission fluxes in the F128N filter to the F110W filter is less than ~1 $\%$). After subtracting the continuum, we obtain the residual flux of the galaxy $F_{\rm{CIII]}} = 3.34\pm1.81 \times10^{-18}\  \mathrm{erg\ s^{-1}\ cm^{-2}}$. The residual flux is calculated by $F_{\rm{CIII]}} = (F_{\rm{F128N}}-F_{\rm{F125W}})*\rm{FWHM_{F128N}}$. There is a tentative 2-$\rm\sigma$ excess of flux measured in F128N compared to the continuum level. We put a 3-$\rm\sigma$ upper limit of  $F_{\rm{CIII]}} \leq 5.43\times10^{-18}\ \mathrm{erg\ s^{-1}\ cm^{-2}}$ and we place a 3-$\sigma$ upper limit for EW of $6.57\ \rm{\AA}$. We also examine the morphology of this galaxy. From the surface brightness profile in F110W imaging, this galaxy may contain two components. The continuum-subtracted image is shown in Figure \ref{fig:figuresdf}.

The photometric results for the four galaxies in the SXDF are summarized in Table \ref{tab:sxdf}. We do not detect CIII] in any of those galaxies.  Their 3-$\rm\sigma$ upper limits on CIII] EWs range from 4.47 $\mathrm{\AA}$ to 29.70 $\mathrm{\AA}$. Source 2 is the brightest galaxy in the SXDF field. We place a stringent 3-$\rm{\sigma}$ upper limit on the EW of 4.47 $\mathrm{\AA}$ and a 3-$\sigma$ upper limit of flux of $3.36\times10^{-18}\ \mathrm{erg\ s^{-1}\ cm^{-2}}$. Sources 3 -- 5 reside in the regions which are close to the detector edges. These regions have relatively higher noise levels which result in weaker constraints on CIII] emission. For example, for source 5, we place a 3-$\rm\sigma$ upper limit on the EW of 29.70 $\mathrm{\AA}$, much higher than that of source 2. Further, we stack all five galaxies in both the SDF and SXDF fields \footnote{We stack the F128N and the F125W images using a weighted average (each source weighted by $\rm{\frac {1}{\sigma^{2}}}$ after normalizing the fluxes). For source one, we use the F110W and F160W images to get an averaged F125W image}. We obtain a 3-$\rm\sigma$ upper limit of 4.20 $\mathrm{\AA}$ on the mean CIII] EW and place a 3-$\rm\sigma$ upper limit of $F_{\rm{stack}} \leq 2.55\times10^{-18}\ \mathrm{erg\ s^{-1}\ cm^{-2}}$ on mean flux. In Table \ref{tab:property}, we summarize the constraints of the CIII] emission in our galaxy sample. For a comparison, we further list some previous measurements in galaxies at similar redshifts of $z\gtrsim6$.


\section{DISCUSSION} \label{sec:floats}

 
We have examined the strength of CIII] emission in a sample of five \textit{z} $=5.70$ galaxies. We do not detect CIII] in the galaxies in our sample but report the measurement on CIII] flux ($3.34\pm1.81 \times 10^{-18}\ \mathrm{erg\ s^{-1}\ cm^{-2}}$) for the brightest source in our sample. We place a 3-$\rm\sigma$ upper limit of 4.20 $\mathrm{\AA}$ on the stacked rest-frame EW for all the five galaxies in our sample. 
In the following discussion, we compare our sample galaxies with previous studies \cite[e.g.,][]{Stark2014,Stark2015,Stark2017,Rigby2015} through two main aspects:  (1) galaxy properties (2) sample selection and depth of the survey.
 
\subsection{Galaxy properties}

According to previous studies on CIII] emitters \cite[e.g.,][]{Stark2014,Stark2015,Stark2017,Rigby2015}, several galaxy properties correlate with the strength of CIII] emission. These galaxy properties include galaxy metallicity, galaxy luminosity, Ly$\alpha$ line strength, stellar population, and the ionization parameter. Using our current data, we compare: (1) galaxy luminosity, (2) Ly${\alpha}$ emission, and (3) metallicity, to other galaxies having CIII] detections or constraints in previous literatures. 

In \cite{Stark2014}, intrinsically fainter galaxies tend to have stronger CIII] emission at  \textit{z} $\sim$ 2. 11 out of 14 of strong CIII] emitters that have rest-frame EW $>$ 5 $\mathrm{\AA}$ have a low intrinsic luminosity ($\rm{M_{UV} > -19.30}$) in \cite{Stark2014,Stark2015,Stark2017}. The galaxies in our sample have a higher luminosity ($\rm{M_{UV}}$ ranging from $\sim$ -20.48 to -22.17). If galaxies at \textit{z} = 5.7 have similar properties with those in \cite{Stark2014} at \textit{z} $\approx$ 2, then we expect moderate CIII] EWs ($<$ 5 $\mathrm{\AA}$) for our galaxy sample.

 Previous studies on CIII] emission in luminous galaxies with $\rm{M_{UV}}$ $<-20$ report relatively weaker CIII] EWs. 
\citet{Rigby2015} study the CIII] emission in a sample of galaxies with high luminosity ($\rm{M_{G} < -20}$). They find a large fraction of 17 $\textit{z} \sim$ 2 - 4 gravitational-lensed star forming galaxies and 46 local galaxies have moderate CIII] emission (EWs less than 5 $\mathrm{\AA}$ at rest frame). The composite spectrum of z $\sim$ 3 Lyman Break Galaxies with luminosity $\rm{M_{R} \sim -21}$ in \cite{Shapley2003} and z $\sim$ 2.4 star forming galaxies with luminosity $\rm{M_{R} \sim -21}$ in \cite{Steidel2016} also have EWs of CIII] emissions $\sim 2\ \mathrm{\AA}$ at rest frame. Our galaxy sample has luminosities similar to that in \cite{Rigby2015,Shapley2003,Steidel2016}, and our CIII] constraints are consistent with their results. Such consistency may be due to the similarity in galaxy luminosities. 

 A positive correlation between CIII] EW and Ly$\alpha$ EW is suggested by \cite{Shapley2003} and is further illustrated in \cite{Stark2014,Stark2015} using galaxies at \textit{z} $\sim$ 2 -- 7.  Using the Ly$\alpha$-CIII] correlation determined in these studies, we can estimate the expected CIII] emission for our galaxy sample. For galaxy source~1 with a rest-frame Ly$\alpha$ EW of ${21}^{+3}_{-2}\ \mathrm{\AA}$ \citep{Shimasaku2006}, the rest-frame CIII] EW is expected to be $\approx 1.5\ \mathrm{\AA}$. The 3-$\sigma$ upper limit on CIII] EW of source 1 is $6.57\ \mathrm{\AA}$ (see Table \ref{tab:property}), consistent with the correlation between Ly$\alpha$ and CIII] emission. Using this CIII]-Ly$\alpha$ correlation, we can estimate the CIII] strength for the other galaxies in the SXDF field (source~2 -- source~5).   The galaxies in the SXDF have Ly$\alpha$ EWs (rest-frame) ranging from ${61.6}^{+21.8}_{-13.1}$ $\mathrm{\AA}$ to ${106.4}^{+107.0}_{-40.3}$ $\mathrm{\AA}$. If we assume the rest-frame Ly$\alpha$ EWs range from ${61.6}$ $\mathrm{\AA}$ to ${106.4}$ $\mathrm{\AA}$ in these galaxies, the rest-frame CIII] EWs are expected to be $\sim 4.0\ \mathrm{\AA} - 8.0\ \mathrm{\AA}$. This expected EW range is lower than our current 3-$\sigma$ upper limits on EWs in these sources (15.09 $\mathrm{\AA} - 29.70\ \mathrm{\AA}$). Thus, we conclude that our results are generally consistent with the Ly$\alpha$ - CIII] correlation found in previous studies (e.g., \cite{Shapley2003,Stark2014,Stark2015}). 

The CIII] strength is also strongly affected by galaxy metallicity. Strong CIII] emitters require a metal-poor interstellar medium with $Z\approx {0.04-0.13\ Z_{\odot}}$ \citep{Stark2014}.  
Various authors \cite[e.g.,][]{Caldwell1992,Guseva2009,Savaglio2005,Salzer2005,Izotov2011,Izotov2014,Zahid2011,Onodera2016} further suggest that there is a correlation between the galaxy luminosity and galaxy metallicity. 
Based on the luminosity and metallicity relation (L-Z relation) fitted by \cite{Zahid2011} at $\textit{z}$ $\sim$ 0.8, the galaxy magnitudes of $\rm{M_{UV} < -20.00}$ correspond to metallicities of  $\rm{12 + log(O/H) > 8.78\pm0.23}$.  
Note the true errors will be much larger than 0.3 dex due to systematical uncertainties in this conversion, and the overall normalization in metallicity in \cite{Zahid2011} tends to bias the estimated metallicity to a larger value. We also need to take into account that galaxies of fixed luminosity will have lower metallicity at higher redshift. Through this conversion, the higher galaxy luminosities in our sample, with $\rm{M_{UV}}$ ranging from $\sim$ -20.48 to -22.17, may correspond to a significantly higher metallicity (e.g., $>$ 0.1 $Z_\odot$) compared with the strongly-lensed galaxy sample in \cite{Stark2014,Stark2015}$ (\rm{M_{UV} > -19.30}$). Since CIII] emission is sensitive to the metallicity \citep{Stark2014}, relatively higher metallicilty in our sample may yield lower EWs compared to galaxies in \cite{Stark2014,Stark2015}. 

\subsection{Sample selection}

\citet{Stark2017} search for the CIII] emission in a sample of two \textit{z} $\ge$ 7 luminous galaxies by using deep ground-based spectroscopy. From the photometry in rest-frame optical, these two galaxies are expected to have strong [OIII] and H$\beta$ emission.  The spectroscopic observations on these two galaxies suggest that the galaxy at $z=7.73$ has a robust detection on CIII] \citep{Stark2017}, with a CIII] EW of $\mathrm{22\pm2\ \AA}$. The other galaxy has a 3-$\sigma$ upper limit  of 7.1 $\mathrm{\AA}$ on the CIII] EW \citep{Stark2017}. The galaxy sample we selected is only based on their redshifts, without any consideration of the strength of [OIII] and H$\beta$ emission. Thus, compared with the extremely strong CIII] emitters in \citet{Stark2017}, our selection technique is different, which could be another reason for the moderate CIII] EWs. 

 Also, our survey is slightly shallower than the extremely deep spectroscopy presented in \cite{Stark2015,Stark2017}. The 1-$\sigma$ flux limit  is $0.4\times10^{-18}\ \mathrm{erg\ s^{-1}\ cm^{-2}}$ in \cite{Stark2015}, smaller than our depth of $ {1\sigma \sim 1\times10^{-18}\ \mathrm{erg\ s^{-1}\ cm^{-2}}}$. If we assume source~1 indeed has a CIII] line flux of $3.34\times10^{-18}\ \mathrm{erg\ s^{-1}\ cm^{-2}}$ (see Table \ref{tab:sxdf}), then we should be able to detect it with a 3-$\sigma$ level with the depth in \cite{Stark2015}. Deeper data may be needed to fully characterize the CIII] emission in source 1, the brightest LAE at $z=5.7$. 

\section{CONCLUSIONS}

Using twelve-orbits of $\it{HST}$/F128N narrowband imaging and two-orbits $\it{HST}$/F125W broadband imaging, we investigated the CIII] emission in a sample of 5 galaxies in the SDF and the SXDF. 
We report a non-detection on CIII] emission for the galaxies in our sample. Using the stacked image, we place a 3-$\rm\sigma$ upper limit of $4.20\ \mathrm{\AA}$ on the mean EW of 5 galaxies. Our photometric results suggest that the CIII] emission may be moderate in a relatively high luminosity LAE sample at $z$ = 5.7, with galaxy magnitudes ranging from $\sim$ -20.48 to -22.17. \cite{Stark2017} suggest strong CIII] emissions may also be found in galaxies with high intrinsic luminosity and low metallicity.  Current samples on CIII] candidates are biased towards luminous spectroscopically confirmed galaxies at z $>$ 5.7. Targeting galaxies with fainter intrinsic luminosities is needed in the future for fully understanding the rest-frame UV/optical emission lines in galaxies in or at the end of the re-ionization epoch. Also,  in five candidates of CIII] emitters of spectroscopically confirmed galaxies at $z$ $> $ 6 \citep{Stark2015,Stark2017,Watson2015}, three of them are found to be CIII] emitters with EWs ranging from $\sim$ 7.6 $\mathrm{\AA}$ to 22.5  $\mathrm{\AA}$ \citep{Stark2015,Stark2017}, while two of them are non-detections \citep{Stark2017,Watson2015}. The present studies (this work and \cite{Stark2015,Stark2017,Watson2015}) on CIII] emission suggest that the CIII] EWs at $z$ $>$ 5.70 have a wide range of distribution. Future facilities, including Giant Segmented Mirror Telescopes (GSMT) and the James Webb Space Telescope (JWST), will thoroughly probe the CIII] emission in a much fainter high-$z$ galaxy population and measure other rest-frame UV/optical emission lines to fully characterize the properties of the typical galaxies at the reionization epoch. 
\acknowledgments
We thank the support from NASA through grant HST-GO-13644 from the Space Telescope Science Institute. Support for part of this work was also provided by NASA through the Hubble Fellowship grant HST-HF2-51370 awarded by the Space Telescope Science Institute, which is operated by the Association of Universities for Research in Astronomy, Inc., for NASA, under contract NAS 5-26555. Daniel P. Stark acknowledges support from the National Science Foundation through the grant AST-1410155.

\newpage

\floattable
\begin{deluxetable*}{cccccc}
\tablecaption{Photometry Results for Five Galaxies in Our Sample\label{tab:sxdf}}
\tablecolumns{7}
\tablenum{1}
\tablewidth{0pt}
\tablehead{
\colhead{No.} &
\colhead{R.A} &
\colhead{Decl.} &
\colhead{F128N Flux Density} &
\colhead{F125W Flux Density} &
\colhead{CIII] line fluxes} \\
\colhead{} &
\colhead{(J2000.0)} &
\colhead{(J2000.0)} &
\colhead{($10^{-20}\mathrm{erg\ s^{-1}\ cm^{-2} \AA^{-1}}$)} &
\colhead{($10^{-20}\mathrm{erg\ s^{-1} cm^{-2} \AA^{-1}}$)} &
\colhead{($10^{-18}\mathrm{erg\ s^{-1} cm^{-2}}$)} 
}
\startdata
1 & 13:24:16.13 & 27:44:11.62 &   $14.30\pm1.10$  & $12.20\pm0.30$ &  $3.34\pm1.81$\\ 
2 &02:17:47.32 &-05:26:48.0  &   $10.60\pm0.70$ &$ 11.30\pm0.10$  & $-1.11\pm1.12$ \\
3 &02:17:45.03 &-05:28:42.5  & $3.50\pm0.68$  & $3.28\pm0.12$  &$ 0.35\pm1.10$  \\   
4 & 02:17:50.00 &-05:27:08.2 &   $1.42\pm0.64$ & $2.31\pm0.13$   & $-1.42\pm1.04$ \\ 
5 &02:17:49.13 &-05:28:54.2 &  $2.19\pm0.81$  & $2.00\pm0.15$  & $0.30\pm1.31$  \\
Stacking & - & -& $8.64\pm0.53$&$8.96\pm0.08$ & $-0.51\pm0.85$\\
\enddata
\tablecomments{The coefficients of the aperture for source 1 to 5 is a kron factor of 1.8 and a minimum aperture of 2.5 semi-major radius.}
\tablecomments{The F125W flux of source 1 is from the UV-continuum fitted by photometry results in F110W and F160W}
\end{deluxetable*}
\vspace{0mm}


\floattable
\begin{deluxetable*}{ccccccc}
\tablecaption{Properties of Galaxies in Our work and Previous Work\label{tab:property}}
\tablecolumns{7}
\tablenum{2}
\tablewidth{0pt}
\tablehead{ 
\colhead{Source} &
\colhead{$\rm{\textit{z}_{Ly{\alpha}}}$} &
\colhead{$\rm{M_{UV}}$} &
\colhead{Ly${\alpha}$ EW} &
\colhead{CIII] EW} &
\colhead{CIII] flux or upper limits} &
\colhead{Reference} \\
\colhead{} &
\colhead{} &
\colhead{} &
\colhead{($\mathrm{\AA}$)} &
\colhead{($\mathrm{\AA}$)}&
\colhead{($10^{-18}\mathrm{erg\ s^{-1}\ cm^{-2}}$)} &
\colhead{} 
}
\startdata
1& 5.70 &$-22.17$ & ${21}^{+3}_{-2}$& $ \leq6.57(3\rm{\sigma})$ & $ \leq5.43(3\rm{\sigma})$  & this work, (1)\\ 
2& 5.69 &$-22.20 $ & - & $\leq4.47(3\rm{\sigma})$& $\leq3.36(3\rm{\sigma})$ & this work, (2)\\
3& 5.75 &$ -20.86$ &${61.6}^{+21.8}_{-13.1}$&$ \leq15.09(3\rm{\sigma})$ & $\leq3.30(3\rm{\sigma})$ & this work, (2), (3)\\
4& 5.69 &$-20.54$ &${106.4}^{+107.0}_{-40.3}$&$ \leq19.95(3{\sigma}) $& $\leq3.12(3\rm{\sigma})$ & this work, (2), (3) \\ 
5& 5.70 &$-20.48$ &${79.3}^{+22.1}_{-14.2}$&$\leq29.70(3\rm{\sigma})$ & $\leq3.93(3\rm{\sigma})$ & this work, (2), (3)\\  
Stacking & - & - & - & $ \leq4.20(3\rm{\sigma}) $ & $ \leq2.55(3\rm{\sigma}) $ & this work \\     \hline \hline 
EGS-zs8-1& 7.73 & $-22.23$& $21\pm4$ & $22\pm2$ & $8.1\pm0.71$ & (4) \\
EGS-zs8-2 &7.48 &$-22.08$&$9.3\pm1.4$& $< 7.1(3{\sigma})$ & $< 2.3(3\rm{\sigma})$ & (4) \\ 
GN-108036 & 7.21 &  $ -21.89$ & 33 &$ 7.6\pm2.8 $ & $2.2\pm0.8$ & (5)\\
 A383-5.2& 6.03 &  $-21.57$& 138 & $22.5\pm7.1$ & $8.9\pm2.7$ & (5)\\
 A1689-zD1& 7.50 &$-22.11$ & - &$\le4(3\rm{\sigma})$ & $\le2(3\rm{\sigma})$ & (6)\\  
\enddata
\tablecomments{Top five sources are galaxies in our work and bottom five sources are galaxies in \cite{Stark2015,Stark2017,Watson2015}}
\tablecomments{EW represents Equivalent Width (rest frame). All upper limits are 3-$\rm{\sigma}$ upper limits.}
\tablecomments{$\rm{M_{UV}}$ for source 2 - 5 and bottom five sources is calculated by assuming a average UV-slope = -2.3 found in a large sample of \textit{z} $>$ 6 galaxies \citep{Jiang2013}}
\tablecomments{Reference: (1): \cite{Shimasaku2006} (2): \cite{Ouchi2005} (3): \cite{Ouchi2008} (4): \cite{Stark2017} (5): \cite{Stark2015} (6): \cite{Watson2015}}
\end{deluxetable*}
\vspace{0mm}

\bibliographystyle{aasjournal.bst}




\listofchanges

\end{document}